\magnification=1200
\tolerance=10000
\hsize 14.5truecm
\hoffset 1.25truecm
\font\cub=cmbx12
\font\ninerm=cmr9
\baselineskip=24truept
\parindent=1.truecm
\def\ref{\par\noindent\hangindent 20pt}

\def\mincir{\raise -2.truept\hbox{\rlap{\hbox{$\sim$}}\raise5.truept
\hbox{$<$}\ }}
\def\magcir{\raise -2.truept\hbox{\rlap{\hbox{$\sim$}}\raise5.truept
\hbox{$>$}\ }}
\def\asymp{\raise -4.3truept\hbox{$ \ \widetilde{\phantom{xy}} \ $}}

\def\di{\displaystyle}
\def\dddot#1{\hbox{\rlap{\raise 8.8truept\hbox{ ...}}
\hbox{$#1$}\ }}
\input newsym
 
\centerline{{\cub GRAVITATIONAL WAVE INTERACTION}}
\medskip
\centerline{{\cub WITH NORMAL AND SUPERCONDUCTING CIRCUITS}}
\vskip 1.5truecm
\centerline{{\bf Pierluigi Fortini},\footnote {$^*$} {\baselineskip=14
truept {\ninerm Department of
Physics, University of Ferrara and INFN Sezione di Ferrara,
I-44100 Ferrara, Italy.}}
 ~{\bf Enrico Montanari$^*$}, 
~{\bf Antonello Ortolan}\footnote {$^\#$}
{\baselineskip=14truept{\ninerm INFN National Laboratories of Legnaro, 
I-35020 Legnaro, Padova, Italy.}}}
\centerline{and}
\centerline{{\bf Gerhard Sch\"afer}\footnote {$^\diamondsuit$}
{\baselineskip=14truept{\ninerm Max-Planck-Gesellschaft,
AG Gravitationstheorie an der Friedrich-Schiller-Universit\"at,
D-07743 Jena, Germany.}} }
\vskip 1.5truecm
\noindent
{\bf Abstract} The interaction, in the long--wavelength approximation,
of normal and superconducting electromagnetic circuits with gravitational
waves is investigated. We show that such interaction takes place by 
modifying the physical parameters R, L, C of the electromagnetic devices. 
Exploiting this peculiarity of the gravitational field we
find that a circuit with two plane and statically charged condensers set at
right angles can be of interest as a detector of periodic gravitational waves.
 
\vskip 2truecm
\centerline{Annals of Physics {\bf 248}, 34--59 (1996)}
\vfill\eject

\noindent
{\bf 1. Introduction}
\bigskip
Future gravitational--wave astronomy requires gravitational--wave detectors
of fantastic high sensitivity. Two designs of detectors are on the beam
of realisation: bar detectors and laser--beam detectors. Both
are promising
to become enough sensitive for the detection and even for the detailed
measurement
of cosmic gravitational waves. Whereas the beam detectors will reach
rather high
sensitivity the bar detectors will be quite cheap. The sensitivity of
the latter should be sufficient to detect gravitational waves from
coalescing binaries in the VIRGO cluster and from type--II supernovae
in our galaxy. Unfortunately, both of them happen not very often (about one
per 30 years).
However, there might exist other sources in our galaxy,
like slightly deformed rotating
neutron stars, in particular young asymmetric pulsars, or even completely
unexpected sources which make gravitational--wave surveys
in our galaxy, or other galaxies of the Local Group, very desirable.
The centers of galaxies are very difficult to investigate with present--day
telescopes so that our knowledge of the centers is very crude.
Gravitational--wave detection might reveal interesting objects there.
 
Purpose of this paper is to work out in details a general formalism for 
treating the interaction of a gravitational wave with an 
electromagnetic (ohmic and superconducting) device in the 
quasi--stationary approximation. Such a formalism is in fact lacking 
and only partial cases are to be found in the literature (see ref. 
[1,3,11,16]).
 
As an application of this general theory we introduce and investigate 
a new design of detectors, normal and superconducting electromagnetic 
circuits which couple to gravitational waves in {\it non--parametric} 
manner. These devices are circuit--generalisations of a capacity--device 
approach invented by Mours and Yvert in 1989. As we shall see in the 
following our devices could reach sensitivities comparable with the 
sensitivities of mechanical detectors for periodic radiation 
[20,21].
 
The paper is organized as follows. In Sect. 2 we discuss,
in quasi--stationary approximation, the interaction
of electromagnetic fields and currents with
weak external gravitational fields. In particular we discuss under
which conditions the conductors, i.e. the ions of the lattices
of the conductors, can be treated as freely falling in external
gravitational wave fields. Sect. 3 presents the general theory of
normal conducting circuits in the field of external gravitational waves.
In this section the parametric nature of the interaction is shown.
Sect. 4 refines the results of Sect. 3 to superconducting circuits.
In Sect. 5 the influence of dielectric and magnetic matter is investigated.
Finally, in Sect. 6, we apply our theoretical results to concrete, 
thermal--noise--dominated
detectors and derive sensitivity--limits for the detection of gravitational
waves.
 
\bigskip
\noindent
{\bf 2. Electromagnetic Fields and Conductors in Curved Spacetime}
\medskip
 
The electromagnetic field equations and the equations of motion for charged
matter (ions and electrons) in an external
gravitational field $g_{\mu\nu} (\mu, \nu = 0,1,2,3)$ can be derived from
the action
$$
S= S_{em} +S_m +S_{int},
\eqno(2.1)
$$
where $S_{em}=$ $(1/4\pi c)\int d^4x \sqrt{-g}\ F_{
\mu\nu} F^{\mu\nu}$ ($g={\rm det} (g_{\mu\nu})$) is the action for the free
electromagnetic field, $S_m$ is the action for the conductors
that we will discuss below, and $S_{int}=\di{{1\over c^2}}\int
d^4x \sqrt{-g} \ j^\mu A_\mu$ , with the current density defined by
$$
j^\mu = {c\rho\over \sqrt{-g_{00}}} {dx^\mu\over dx^0},
$$
is the action
for the coupling between the
electromagnetic field and the charges and currents in the conductor.
The electromagnetic field--strength tensor $F_{\mu\nu}$ is related with the
four--vector potential $A_\mu$ by $F_{\mu\nu}=A_{\nu,\mu}-A_{\mu,\nu}$ and
we have the following relation between
$F_{\mu\nu}$ and the standard electromagnetic field--strength entities, $E_i$
and $B_i$ ($i = 1,2,3$),
$$
F_{\mu \nu} =
\left(
\matrix {0 & -E_1 & -E_2 & -E_3 \cr
E_1 & 0 & B_3 & -B_2 \cr
E_2 & -B_3 & 0 & B_1 \cr
E_3 & B_2 & -B_1 & 0}
\right).
\eqno(2.2)
$$
 
By varying the action (2.1) with respect to $A_\mu$ we get the De Rahm
equations. In the Lorentz gauge, $\nabla_\nu A^\nu =0$, they read
 
$$
\nabla^\nu\nabla_\nu A^\mu - A^\rho R_\rho^{ \ \mu}=
-{4 \pi \over c}\ j^{\mu}.
\eqno(2.3)
$$
Eq. (2.3) generalizes the inhomogeneous Maxwell equations in the
Lorentz gauge to curved spacetime;
$R_{\mu\nu}$ is the Ricci tensor defined by
$$
R_{\mu \nu} = R^\alpha_{\ \  \mu \alpha \nu},
$$
where $R^\alpha_{\ \ \mu \beta \nu}$ is the Riemann tensor
$$
R^\alpha_{\ \ \mu \beta \nu} = {\partial \Gamma ^\alpha _{\mu \nu} \over
\partial x^\beta} - {\partial \Gamma ^\alpha _{\mu \beta} \over
\partial x^\nu} + \Gamma^\alpha_{\eta\beta} \Gamma^\eta_{\mu\nu} -
\Gamma^\alpha_{\eta\nu} \Gamma^\eta_{\mu\beta},
\eqno(2.4)
$$
and where $\Gamma^\alpha_{\mu\nu}$ denote the Christoffel symbols
$$
\Gamma ^\alpha _{\mu \nu} = {1\over 2} g^{\alpha \beta}
\biggl( {\partial g_{\beta \nu} \over \partial x^\mu} +
{\partial g_{\beta \mu} \over \partial x^\nu} -
{\partial g_{\mu \nu} \over \partial x^\beta}\biggr),
\eqno(2.5)
$$
which enter in the definition of the covariant derivative $\nabla_\mu$
(notations and conventions as in ref. [4] pp. 223--4).
 
Our aim is to study the interaction of a gravitational wave, emitted
for instance by a cosmic source, with electromagnetic circuits.
In the interaction region the metric tensor $g_{\mu\nu}$ satisfies
the Einstein equations in vacuo, $R_{\mu \nu} = 0$.
For weak gravitational fields, to which we shall limit ourselves in
this paper, the metric tensor and its determinant can be written as
$$
\eqalign{
&g_{\mu \nu} = \eta_{\mu \nu} + h_{\mu \nu},\cr
&\sqrt{- g}=1+{h \over 2},\cr}
\eqno(2.6)
$$
where $\eta_{\mu \nu} = {\rm diag}(-1,1,1,1)$ is the Minkowski metric tensor
and where $h_{\mu \nu}$, with $|h_{\mu \nu}|<<1$, is its small perturbation;
$h\equiv h^\mu_{\ \mu}
=\eta^{\mu\nu} h_{\mu\nu}$. In this approximation we can fix a
particular coordinate frame, respectively gauge, the so--called TT 
(transverse and
traceless) gauge [22] specified by $h=0$, 
$h_{0 \mu}=0$,
and $\partial_j h^{ij}=0$, where the solutions $h_{\mu\nu}$ of the
homogeneous Einstein equations take the very simple form:
$$
h_{ij}= A_+ ({\bmit e}_+)_{ij}
e^{i (k_\mu x^{\mu} +\varphi_+}) +
A_\times ({\bmit e}_\times)_{ij}
e^{i (k_\mu x^{\mu} +\varphi_\times}),
\eqno (2.7)
$$
where $k^\mu\equiv (\omega_g/c,{\bmit k})$ and $k^\mu k_\mu=0$;
$A_{+,\times}$ are the amplitudes of the
two polarization states and ${\bmit e}_{+,\times}$ are the two transverse
($e_{ij}k^j=0$) and traceless ($e^i_{\ i}=0$) polarization
tensors [22].
 
Neglecting second order terms in $h_{\mu\nu}$ one can easily show that
the linearized De Rahm equations are
$$
\eqalign {
A^{\mu,\nu}_{\ \ ,\nu} + &h^\mu_{\ \nu,\beta}\ A^{\beta,\nu}
+ h^\mu_{\ \beta,\nu}\ A^{\beta,\nu} - h_{\nu\beta}^{\ \ ,\mu}
\ A^{\beta,\nu} - h^{\nu\alpha}\ A^{\mu}_{\ ,\alpha,\nu} =
- {4 \pi \over c} \ j^{\mu},
\cr 
\partial_\nu \ A^\nu &=0.
\cr}
\eqno(2.9)
$$
We shall consider our electromagnetic system in the so--called
long--wavelength (or quasi--stationary) approximation in which the
frequency of the electromagnetic oscillations $\nu_{em}$ is much smaller
than $c/d$ where $d$ is the typical linear dimension of the system.
This condition is equivalent to the assumption that all time
derivatives of the potential in (2.9) are negligible with respect to
space derivatives and also that the total electromagnetic radiation of
the system is negligibly small (this implies also $v/c<<1$ where $v$
is the typical velocity of the charged particles in the system). If these
conditions are satisfied it has been shown ([9] and e.g. [14] \S 75)
that $S_{em}=-(1/2) S_{int}$ and thus, the total action reduces to
$$
S={1\over 2c^2}\int \sqrt {-g}\ d^4x \ j^\mu A_\mu + S_{m}.
\eqno(2.10)
$$
In order to get a consistent picture of the dynamics of a non--radiating
system we shall request that the reduced gravitational wavelength 
$\lambda_g/2\pi$ is
much greater than $d$. This entails that the leading term in
eq. (2.9) is the one proportional to $h_{ij}$. Within these
approximations we get
$$
A^{\mu,k}_{\ \ ,k} - h^{ij} \ A^\mu_{\ ,i,j} = - {4\pi\over c}\ j^{\mu}.
\eqno(2.11)
$$
 
We shall now investigate the matter action $S_m$.
We shall show that if the elastic bodies forming the circuits can be 
treated as free falling in
the field of the  gravitational wave, we can discard the
mechanical part of the circuit.
It is a well known (see [22]) property of the TT system that if
freely falling bodies are at rest before the arrival of the wave
their spatial TT coordinates remain constant in time also in the
oscillating gravitational field. This fact will greatly simplify our
treatment as all the relevant circuit parameter perturbations
can be found in TT coordinates by integration over unperturbed 
coordinate paths.
 
To find the free-falling conditions let us consider the
normal modes of a solid body as a collection of non--coupled damped 
harmonic oscillators.
This problem can be easily solved in the Fermi Normal Coordinates 
(FNC) frame (see [5]). The deformation from the 
equilibrium shape can be written as a sum
$$
\delta \bmit x = \sum_{n,l} A_{nl}(t) \bmit {\psi}_{nl}(\bmit x)
\eqno(2.12)
$$
The $l$--label refers to the kind of modes: for instance, as far as
cylindrical bars are concerned, it refers to the longitudinal, 
torsional and flexural modes. The $n$--label refers to the eigenfrequencies 
of a given mode. 
 
For each specific value of $l$ the fundamental mode shall be denoted
by $n=1$. In the following we assume that only one specific $l$--mode 
is strongly interacting with the gravitational wave. For instance, 
for a spherical body $R_{nl}$ is different from zero only when $l=2$.
 
In the presence of a gravitational wave, the coefficients $A_{nl}$ 
are determined by the equation (see [4] eqs. (37.42b--c)):
$$
\ddot A_{nl} + {1\over \tau_{nl}} \dot A_{nl} + 
\omega^2_{nl} A_{nl} = R_{nl},
\eqno(2.13)
$$
where
$$
R_{nl} = {1\over 2} {1\over M} \ddot h_{ij}
\int \psi^i_{nl}x^j \rho d^3x
\eqno(2.14)
$$
($\rho$ is the mass density of the body and M its total mass).
The solution in Fourier space ($\xi(\omega)=$ 
$(2\pi)^{-1/2}\int_{-\infty}^{+\infty} \xi(t) \exp(-i\omega t) \ dt$)
is given by 
$$
A_{nl} (\omega) = - {1\over 2} t^{ij}_{nl}  h_{ij}(\omega) 
{\omega^2 \over \omega_{nl}^2-\omega^2 +
i\omega/\tau_{nl}},
\eqno(2.15)
$$
where $t^{ij}_{nl}$ is defined as 
$$
t^{ij}_{nl} = {1\over M} \int \psi^i_{nl}x^j \rho d^3x.  
\eqno(2.16)
$$
 
For monochromatic gravitational waves of frequencies 
$\omega_g>>\omega_{1l}$ we get
$$
A_{1l}(t)={1\over 2} \ t^{ij}_{1l} \ h_{ij}(t)
\eqno(2.17)
$$
that is to say, in its fundamental mode the body behaves as a 
freely moving system. 
This is the operating condition for the interferometric
devices under construction in Europe (VIRGO) and in the United States
(LIGO). 
 
Let us now consider the opposite case where $\omega_g<<\omega_{1l}$.
We get
$$
A_{nl}(t)= -{1\over 2}\ t^{ij}_{nl}\ h_{ij}(t) 
\bigg({\omega_g \over \omega_{nl}}\bigg)^2
\eqno(2.18)
$$
that is the body is practically at rest and the geometrical properties
of the circuit do not change.  
 
Finally at mechanical resonance of the fundamental mode 
$\omega_g=\omega_{1l}$ we get
$A_{1l}(\omega)= i t^{ij}_{1l}\ h_{ij}(\omega) {\cal Q}_{1l}$ where 
${\cal Q}_{1l}=\pi\nu_{1l}\tau_{1l}$ is the quality factor of the 
oscillator.
This is the operation condition for the resonant antennae already
working in Italy 
(Universities of Rome and National Laboratories of INFN at Frascati
and Legnaro) and in
the United States (Universities of Louisiana and Stanford).
 
Now, in order to find free falling conditions in our case, we examine 
in more details $R_{nl}$. Comparing eq. (37.45) with eq. (7b) Box 37.4
and taking into account definition (4) Box 37.4 of [4], it can be reasonably 
assumed that $R_{nl}$ may have the following behaviour as a function of $n$
$$
R_{nl} \simeq {1\over n^2} R_{1l}
\eqno(2.19)
$$
This implies that the energy deposited in the n--th normal mode at resonance
is approximately independent on $n$ (see [4] eq. (37.43)) if we assume
$\omega_{nl} \simeq n \omega_{1l}$.
With these assumptions the $A_{nl}$ coefficients are given by
$$ 
A_{nl} (\omega) = -{1\over 2} {1\over n^2} t^{ij}_{1l} h_{ij} (\omega)
{\omega^2 \over 
n^2 \omega_{1l}^2-\omega^2 + i\omega/\tau_{nl}}   
\eqno(2.20)
$$
If we now consider again an incoming gravitational wave with frequency 
$\omega_g>>\omega_{1l}$, different from all the other 
$\omega_{nl}$, then from eq. (2.20) one can see that only the 
fundamental mode is practically excited.
For instance if the first harmonic excited by the wave is $n=3$ 
(see, e.g. [4], eq. (37.45)) then 
$A_{3l} \simeq {1\over 10} A_{1l}$.
Therefore the body is free falling because it follows the 
incoming gravitational wave like a cloud of dust (see eq. (2.17)) and so 
it is practically at rest in TT gauge.
One can see therefore that, by suitably arranging the fundamental mode
frequency $\omega_{1l}$ with respect to the frequency chosen for
observing gravitational radiation, the condition of free falling can be 
easily fulfilled.
In the rest of this paper we shall restrict ourselves to physical devices
where these conditions are satisfied and therefore we neglect
the matter--action term in (2.1). 
 
\bigskip
\noindent
{\bf 3. General Theory of Electromagnetic Circuits}
\medskip
 
To establish the behaviour of a circuit in the field of a weak
gravitational wave we generalize the lagrangian approach
of [15].
Let $\gamma_a$ ($a=1,2,\dots,N$) be a system of $N$ conducting
non--ferromagnetic one dimensional (wires) or extended bodies
(capacitors). The quasi--stationary approximation allows us to
neglect surface currents in extended conductors and
charged wires. The charge over conductor surface and
the current flowing in wires can be treated as not depending on spatial
coordinates. The relations between, on the one side,
$j^\mu$ and the current $I$ and, on the other side, charge density $\rho$
and the charge $dQ$ in an infinitesimal volume element can be
written in any coordinate system as (see for instance [14] \S 90)
$$
\cases{&$
Idx^i=j^i\ \sqrt{-g}\ d^3x
$\cr&
\cr&$
dQ=\rho\ \sqrt{\ ^3\!g}\ d^3x
$\cr},
\eqno(3.1)
$$
where $^3\!g$ is the determinant of the three--dimensional metric
$\ ^3\!g_{ij}=g_{ij}-g_{0i}g_{0j}/g_{00}$.
The equations for
$I$ and $Q$ can be found by substituting in the action (2.10) the (formal)
solution of the De Rahm equation (2.3) for gravitational waves. If we can 
neglet $g_{\mu\nu,\alpha}$ with respect to $g_{\mu\nu}$ (i.e. for slowly 
varying gravitational fields) then 
$\quadratello \equiv \nabla^\mu\nabla_\mu \approx
g^{\alpha\beta}\partial_\alpha\partial_\beta$. In this
case we can write
$$
A_\mu(x) ={1\over c}
\int_{\Omega}{\overline K_{\mu\nu}
[x,x']
j^\nu(x')\sqrt{-g(x')}\ d^4x'},
\eqno(3.2)
$$
where ${\overline K_{\mu\nu}}$ is the retarded Green function,
solution of the equation
$$
\quadratello_x\ \overline K_{\mu\nu}
[x,x']=
-4\pi {g_{\mu\nu}(x')\over\sqrt {-g(x')}}
\delta^4(x^\alpha-x'^\alpha)
\eqno (3.3)
$$
In a synchronous reference frame $x\equiv (ct,{\bmit x)}$,
where $\ ^3\!g_{ij}=g_{ij}$ and $-g=\ ^3\!g$, we can integrate eq. (3.2)
over the time coordinate $x'^0$ and get the generalized retarded
potential ([4] p. 500) 
$$
A_\mu({\bmit x},t) ={1\over c}
\int_{V}{K_{\mu\nu}[x,\bmit x']
j^\nu(x^0,\bmit x')
\sqrt{\ ^3\!g(x^0,\bmit x')}d^3x'}.
\eqno(3.4)
$$
Now we have to introduce the lagrangian ${\cal L}$ of a system. 
>From (2.10) and (3.4) we obtain the total
lagrangian of a system of conductors and wires in curved
spacetime
$$
\eqalign {
{\cal L}={1\over 2c^2}\int_V{\sqrt{\ ^3\!g(x)}\ d^3x}
\int_{V}&{\sqrt{\ ^3\!g(x^0,\bmit x')}\ d^3x'} \times
\cr
\times
\ &K_{\mu\nu}[x,\bmit x']
\ j^\mu(x)\ j^\nu(x^0,\bmit x')
\cr}
\eqno(3.5)
$$
>From the relations (3.1), taking into account that $I$ and $Q$ are only
functions of time in the
quasi--stationary--field
approximation and that $\overline K^{0i}=0$ in synchronous reference
frames, we get
$$\eqalign{
{\cal L}&= \sum_{a,b=1}^{N} {I_{\gamma_a} \ I_{\gamma_b} \over
2c^2}\int_{\gamma_a}\!dx^i\int_{\gamma_b}\!dx'^j
K_{ij}[x,\bmit x']
\cr
&+{1\over 2}\int_{\gamma_a}\!dQ\int_{\gamma_b} \!dQ'
K_{00}[x,\bmit x'].
\cr}
\eqno(3.6)
$$
For quasi--stationary fields charges are distributed over the conductor
surfaces $A_{\gamma_a}$ and so we can replace in (3.6)
the charge integrals with surface
integrals. This leads to
$$
{\cal L} =  \sum_{a,b=1}^N \biggl[ {1\over 2}\ 
L_{\gamma_a\gamma_b}(t) I_{\gamma_a} \ I_{\gamma_b} -
{1\over 2} \ 
{\widehat C_{\gamma_a\gamma_b}}(t) Q_{\gamma_a} Q_{\gamma_b}\bigg]\ ,
\eqno(3.7)
$$
where $L_{\gamma_a\gamma_b}$ are the generalized coefficients of mutual
inductance
$$
L_{\gamma_a\gamma_b}(t)=
{1\over c^2}\int_{\gamma_a}{dx^i}\int_{\gamma_b}{dx'^j}
K_{ij}[x,\bmit x'],
\eqno(3.8)
$$
and $\widehat C_{\gamma_a\gamma_b}$
$$
Q_{\gamma_a}{\widehat C_{\gamma_a\gamma_b}}(t)Q_{\gamma_b}=
-\int_{\gamma_a}{dA_{\gamma_a}}\int_{\gamma_b}{dA'_{\gamma_b}}
{dQ_{\gamma_a} \over d A_{\gamma_a}}
K_{00}[x,\bmit x']
{dQ_{\gamma_b} \over d A'_{\gamma_b}},
\eqno(3.9)
$$
define univocally the coefficients of capacity $C_{\gamma_a\gamma_b}$ 
by means of the equations 
$\sum_{b=1}^N \widehat C_{\gamma_a\gamma_b} C_{\gamma_b\gamma_c}=$
$\delta_{ac}$.
>From the above formulae one easily recognizes that the action of
gravitational fields in synchronous reference frames (such as TT--gauge frames)
is a change in the geometrical circuit parameters, i.e. the interaction of
gravitational radiation with standard electromagnetic circuits is of
parametric nature.
 
In order to take into account also dissipations of real circuits we go
over to Ohm's law written in its covariant form ([8] p. 263);
for the current in wires one has
$$
j^\alpha - j_\beta \ u^\beta \ u^\alpha =
\sigma \ F^\alpha_{\ \nu} \ u^\nu\ ,
\eqno(3.10)
$$
where $u^\nu$ is the normalized four velocity of the lattice of the conductor
$\gamma_a$, and $\sigma_a$ is its 
conductivity
which we assume constant in the range of frequencies we are 
interested in.
As the lattice has fixed TT coordinates $u^\mu=\delta^\mu_{0}$,
the total dissipated power in our circuits by Joule effect is
$$\eqalign{
P & \equiv \int_V \sqrt{\ ^3\!g}\ d^3x\ j^\mu\ F_{\mu\nu} \ u^\nu \cr
& = \sum_a {1\over\sigma_a} \ \int_{\gamma_a}
\sqrt{\ ^3\!g}\ d^3x\ g_{ij} \ j^i(x^k) \ j^j(x^k), }
\eqno(3.11)
$$
which can be rewritten in the form
$$
P=\sum_{a=1}^N R_{\gamma_a}(t) I_{\gamma_a}^2,
\eqno(3.12)
$$
where $R_a(t)$ are generalized resistances
$$
R_{\gamma_a}(t)={g_{ij}\over\sqrt{\ ^3\!g}}\
\int_{\gamma_a}\ {1\over \sigma_{\gamma_a} S_{\gamma_a}}
\ t_{\gamma_a}^j\  dx^i ;
\eqno(3.13)
$$
here $t_{\gamma_a}^i$ is the unit vector 
($\ ^3\!g_{ij} t_{\gamma_a}^i\ t_{\gamma_a}^j=1$) tangent
to the wire $\gamma_a$ and $dS_{\gamma_a}\equiv d^3x_{\gamma_a}/dl$,
($dl^2=g_{ij}dx^idx^j$) is the cross--section of the wire.
 
In order to get the Euler--Lagrange equations from the lagrangian
(3.7) we have to set up a relation among our charges and currents.
>From (3.1) and the charge conservation relation follows
$I_{\gamma_a}=\dot Q_{\gamma_a}$. 
Hereof the equations of motion result in the form
$$
\sum_{b=1}^N \biggl [
L_{\gamma_a\gamma_b}(t)\ \ddot Q_{\gamma_b}+
\dot L_{\gamma_a\gamma_b}(t)\ \dot Q_{\gamma_b}+
\widehat C_{\gamma_a\gamma_b}(t)\ Q_{\gamma_b} \biggr ]=0.
\eqno(3.14)
$$
Once we have defined the dissipation function $\Phi=P/2$
we can substitute the r.h.s. of the equation (3.14) with
$\partial \Phi/\partial \dot Q_{\gamma_a}$ and write  (see [13] \S 1.5)
$$
\sum_{b=1}^N \ L_{\gamma_a\gamma_b}(t)\ddot Q_{\gamma_b}+
\dot L_{\gamma_a\gamma_b}(t)\dot Q_{\gamma_b} +
\widehat C_{\gamma_a\gamma_b}(t)\ Q_{\gamma_b}= 
- R_{\gamma_a}(t)\dot Q_{\gamma_a}.
\eqno(3.15)
$$
These equations describe, in TT coordinates, $RLC$ circuits
in external gravitational radiation fields when dissipations are
taken into account.

In the weak gravitational field approximation
one can easily show (cf. [10]) that
$$
K_{\mu\nu}=
{\eta_{\mu\nu} + h_{\mu\nu}\over
|\bmit x-\bmit x'|}+
{1\over 2} h_{kl} \partial'^k\partial'^l
|\bmit x-\bmit x'| \eta_{\mu\nu}
\eqno(3.16)
$$
satisfies eq. (2.11) through eq. (3.4), and therefore it gives the
right expression of the Green function $K_{\mu\nu}$ in the linear
and quasi--static field approximation and for a TT--metric
perturbation $h_{\mu\nu}(t)$. From eqs. (3.8),
(3.9) and (3.13) it follows that, in the same approximation, the coefficients
$L_{\gamma_a\gamma_b}(t)$, $\widehat C_{\gamma_a\gamma_b}(t)$ and 
$R_{\gamma_a}(t)$ read
$$\eqalign{
L_{\gamma_a\gamma_b}(t) = 
&\ ^0\!L_{\gamma_a\gamma_b} + l_{\gamma_a\gamma_b}(t)
\cr
\widehat C_{\gamma_a\gamma_b}(t)=
&\ ^0\!\widehat C_{\gamma_a\gamma_b} + \widehat c_{\gamma_a\gamma_b}(t)
\cr
R_{\gamma_a}(t)=&\ ^0\!R_{\gamma_a} + r_{\gamma_a}(t)
\cr}
\eqno (3.17)
$$
where $\ ^0\!L_{\gamma_a\gamma_b}$, 
$\ ^0\!\widehat C_{\gamma_a\gamma_b}$, and $\ ^0\!R_{\gamma_a}$ are
respectively the usual inductance, capacitance, and resistance coefficients
in flat spacetime, defined as
$$
\ ^0\!L_{\gamma_a\gamma_b}={1\over c^2}
\int_{\gamma_a}{}\int_{\gamma_b}{}\delta_{ij}
{dx^i_{\gamma_a}dx'^j_{\gamma_b}\over
|\bmit x_{\gamma_a}-\bmit x'_{\gamma_b}|},
\eqno(3.18)
$$
$$
\ ^0\!R_{\gamma_a}=
\delta_{ij}\int_{\gamma_a}{}
{t_{\gamma_a}^it_{\gamma_a}^j\over\sigma_{\gamma_a}}
{dl\over dA_{\gamma_a}}
\eqno (3.19)
$$
$$
Q^{\gamma_a} \ ^0\!\hat C_{\gamma_a\gamma_b}Q^{\gamma_b} = 
\int_{\gamma_a}{}\int_{\gamma_b}{}
{dQ^{\gamma_a}\over dA_{\gamma_a}}
{dA_{\gamma_a}dA'_{\gamma_b}\over
|\bmit x_{\gamma_a}-\bmit x'_{\gamma_b}|}
{dQ^{\gamma_b}\over dA'_{\gamma_b}}.
\eqno(3.20)
$$
Here $l_{\gamma_a\gamma_b}(t)$,
$\widehat c_{\gamma_a\gamma_b}(t)$ and 
$r_{\gamma_a}(t)$ are the time dependent perturbations
induced by the gravitational waves on the circuit parameters which can be 
written as:
$$
\eqalign{
l_{\gamma_a\gamma_b}=&h_{ij}\lambda^{ij}_{\gamma_a\gamma_b}
\cr
\hat c_{\gamma_a\gamma_b}=&h_{ij}\chi^{ij}_{\gamma_a\gamma_b}
\cr
r_{\gamma_a}=&h_{ij}\varrho_{\gamma_a}^{ij}
\cr}
\eqno(3.21)
$$
with
$$
\lambda^{ij}_{\gamma_a\gamma_b}=
{1\over c^2}
\int_{\gamma_a}{}\int_{\gamma_b}{}
{dx^i_{\gamma_a}dx'^j_{\gamma_b}\over
|\bmit x_{\gamma_a}-\bmit x'_{\gamma_b}|}+
{1\over 2c^2}
\int_{\gamma_a}{}\int_{\gamma_b}{\delta_{kl}}
\partial'^i\partial'^j
|\bmit x_{\gamma_a}-\bmit x'_{\gamma_b}|
dx^k_{\gamma_a}dx'^l_{\gamma_b},
\eqno(3.22)
$$
$$
Q^{\gamma_a}\chi^{ij}_{\gamma_a\gamma_b}Q^{\gamma_b}=
{1\over 2}
\int_{\gamma_a}{}\int_{\gamma_b}{}
{dQ^{\gamma_a}\over dA_{\gamma_a}}
{dQ^{\gamma_b}\over dA'_{\gamma_b}}
\partial'^i\partial'^j
|\bmit x_{\gamma_a}-\bmit x'_{\gamma_b}|
dA_{\gamma_a}dA'_{\gamma_b}
\eqno(3.23)
$$
and
$$
\varrho_{\gamma_a}^{ij}=
\int_{\gamma_a}{}
{t_{\gamma_a}^it_{\gamma_a}^j\over\sigma_{\gamma_a}}
{dl\over dS_{\gamma_a}}.
\eqno(3.24)
$$
The coefficients $\lambda^{ij}_{\gamma_a\gamma_b}$, 
$\chi^{ij}_{\gamma_a\gamma_b}$ and
$\varrho_{\gamma_a}^{ij}$, defined in (3.22), (3.23) and (3.24), are pure
geometric quantities which play a similar
r\^ole as the mass-quadrupole tensor
$Q^{ij} = \int_V \rho(\bmit x)(x^i x^j -  (1/3) \delta^{ij} x^k x_k)
\ d^3x$
in the mechanical interaction between a gravitational wave and
a solid body.
 
We notice that the equations that describe the
interaction between gravitational waves and electromagnetic 
circuits are parametric ones. This means that gravity is not acting as
electromotoric force but is changing the circuit parameters.
The effect of the small perturbation of the parameters of the circuit 
is a frequency and amplitude modulation (see for instance [18]).

\bigskip
\noindent
{\bf 4. The superconducting circuit}
\medskip
 
Below a certain temperature $T_t$, called transition
temperature, some metals become superconducting. The most impressive
properties of this state are the sudden drop of the electrical
resistance and the Meissner effect.
F.~London in $1935$ (cfr. [12]) suggested a macroscopic theory of the pure
superconducting state which unified these two experimental facts under
the same theoretical scheme. He supposed that the density current
flowing in the superconductor was a sum of two currents: the
normal $\bmit j_n$ and the supercurrent $\bmit j_s$
$$
\bmit j=\bmit j_s + \bmit j_n .
\eqno (4.1)
$$
The two kinds of current are related to the electromagnetic field
inside the conductor in different ways. The supercurrent satisfies
the two London equations
$$
\eqalign{
curl\ (\Lambda \bmit j_s) &= - {\bmit B\over c}
\cr
{\partial\over\partial t}\ (\Lambda \bmit j_s)&= \bmit E .
\cr}
\eqno (4.2)
$$
Here $\Lambda$ is a constant characteristic of the superconductor (its
order of magnitude is $>10^{-31}\ sec^2$); $\bmit B$ and $\bmit E$ are
the intensities of the magnetic and electric fields.
The normal current is connected with the electric field by Ohm's law
$$
\bmit j_n = \sigma \bmit E
\eqno (4.3)
$$
where $\sigma$ is a continuous function of the temperature and has no
jump at $T=T_t$. The total current $\bmit j$ satisfies the
Maxwell equations and the relation
between the two currents turns out to be
$$
{\partial\over\partial t}\ \bmit j_s =
{1\over \sigma\Lambda}\ \bmit j_n =\beta\ \bmit j_n
\eqno (4.4)
$$
where $\beta$ has the dimension of a frequency
($\beta=1/\sigma\Lambda\approx 10^{12}\ Hz$). If the
electromagnetic field oscillates at a frequency $\omega_{em}$ we
have
$$
|\bmit j_n| = {\omega_{em} \over \beta}\ |\bmit j_s|
\eqno (4.5)
$$
and if $\omega_{em} <<\beta$ then $|\bmit j_n|<<|\bmit j_s|$; in this
way the sudden drop of the resistance is due only to the fact that the
normal current (which is the one that dissipates) becomes practically
zero.
Moreover, from eq. (4.2) and the Maxwell equations follows
$$
\Delta B_i = {1\over\lambda^2}\ B_i
\eqno (4.6)
$$
where
$$
\lambda = c\ \sqrt{{\Lambda\over 4\pi}}\approx 10^{-5}\ cm
\eqno (4.7)
$$
is the so called penetration depth. The regular solutions of the eq. (4.6)
decrease exponentially from the surface of the superconductor to its interior
with typical length of the order of the penetration depth.
Therefore inside large bodies (large compared with the penetration
depth) we have $\bmit B=0$. 
 
The London theory of superconductivity is enough for our purposes. In
fact, as in the preceeding section we didn't need a microscopical
theory of conductivity so now we need only the knowledge of the
macroscopical behaviour of a superconductor that is given by the eqs.
(4.1)--(4.3).
 
In the following we point out the differences from the ohmic case. The
first step is to write down the equation of an oscillating
superconducting RLC circuit in flat spacetime.  
>From the Maxwell equations we get the energy theorem as usual
$$
div {c\over4\pi}(\bmit E\times\bmit B)+
{\partial\over\partial t}{1\over8\pi}(\bmit B^2+\bmit E^2)=
-(\bmit j\bmit E).
\eqno (4.8)
$$
For quasi--stationary conditions the first term is negligible.
$(1/8\pi)(\bmit B^2+\bmit E^2)$ is the energy density of
the electromagnetic field and it has the same expression as in the ohmic
case. 
 
The differences come out when we write explicitly the work done by the
electric field (see [12] \S 9),
$$
(\bmit j\bmit E) =
{\partial\over\partial t}
\biggl ( {1\over2}\Lambda \bmit j_s^2 \biggr) +
{1\over\sigma} \bmit j_n^2 .
\eqno (4.9)
$$
The term ${1\over2}\Lambda \bmit j_s^2$ represents a reversible work
(it is in fact the kinetic energy density of the supercurrent) while
$(1/\sigma) \bmit j_n^2$ is the energy density dissipated by
the normal current (Joule effect).
Another difference arises from the fact that if
$\omega_{em} <<\beta$ the current is different from zero only
for a few penetration depths below the surface.
Therefore we need to be careful when we calculate the kinetic
energy and the dissipated power.
In fact, let us approximate the current as
$$
\bmit j(r,t)\approx
\bmit j(t)\ exp\ \biggl[-{(r_0-r)\over\lambda}\biggr]
\eqno (4.10)
$$
where $r_0$ is the radius
of the wire ($r_0>>\lambda $); the power lost as Joule heat is
given then by
$$
P=\int_V{{1\over\sigma}\ \bmit j_n^2\ dV}= R_sI_n^2
\eqno (4.11)
$$
where
$$
I_n\approx 2\pi\lambda r_0\ |\bmit j_n(t)|
\eqno (4.12)
$$
and
$$
R_s=\int_\gamma{{dl\over\sigma\eta}}.
\eqno (4.13)
$$
$\eta$ is the effective section in which the current flows. Is takes the
value
$$
\eta \approx 4\pi\lambda r_0.
\eqno (4.14)
$$
In a wire one has $R_s\approx (r_0/\lambda)\ R$,
where R is the usual resistance.
For the kinetic energy we get
$$
W_{kin}=\int_V{{1\over2}\Lambda\ \bmit j_s^2\ dV} =
{1\over 2\beta}\ R_sI_s^2
\eqno (4.15)
$$
where for the last equality see eq. (4.4); $I_s$ is defined
analogously to $I_n$.
 
Therefore we can write the lagrangian of an RLC superconducting
circuit as ($I=I_s+I_n$, $Q=Q_s+Q_n$)
$$
{\cal L} = {1\over 2}\ \biggl ( L+{R_s\over\beta}\biggr)\ I_s^2 +
LI_sI_n + {1\over2} LI_n^2 -
{1\over 2C}Q_s^2 - {1\over C} Q_sQ_n - {1\over 2C}Q_n^2 .
\eqno (4.16)
$$
The dissipation function reads
$$
\Phi = {1\over 2} R_sI_n^2
\eqno (4.17)
$$
and the relation between $I_n$ and $I_s$ is given by (see eq. (4.4))
$$
\dot I_s = \beta I_n .
\eqno (4.18)
$$
Using the method of Lagrange multipliers (see for instance [13] \S 
2.4) we find
$$
\ddot Q + 2\gamma_s \dot Q_n +\omega_0^2 Q = 0
\eqno (4.19)
$$
where $2\gamma_s=R_s/L$ and $\omega_0^2=1/LC$.
The only difference from the equation of an ohmic circuit is in the
resistance term: in the superconducting case this term is only due to
a part of the current (the normal current).
 
Making use of relation (4.18)
we find for the total current a differential equation of the third 
order in time
$$
\dddot I + \beta (1+2\Gamma)\ddot I +
\omega_0^2 \dot I +\beta \omega_0^2 I = 0
\eqno (4.20)
$$
where $2\Gamma = R_s/\beta L << 1$.
The solution of this equation can be written, up to terms of the first
order in $\Gamma$, as
$$
I(t)=Ae^{-\beta(1+2\Gamma)t} + e^{-\gamma_0t}
[B\cos{(\Omega_0t)} + D\sin{(\Omega_0t)}]
\eqno (4.21)
$$
where
$$
\eqalign {
\Omega_0=&\omega_0(1-\Gamma)
\cr
\gamma_0=&{\Gamma \Omega_0^2\over\beta}
\cr}
\eqno(4.22)
$$
The first term is a sort of opening extracurrent which has its origin
when, after starting the current with an external e.m.f. of frequency
$\omega_{emf}$, we let oscillate the circuit freely. The coefficient
$A$ is practically different from zero only if $\omega_{emf}$ is
very different from $\omega_0$. In this case the current must change
its frequency according to (4.18). The characteristic time of this
term, being of the order of $\beta^{-1}$, is
too short to be taken into account in a macroscopical theory like this.
Therefore we can neglect this term and put $A=0$.
In this case it is easy to recognize that the following equation
$$
\ddot I +2\gamma_0\dot I+\Omega_0^2 I=0
\eqno (4.23)
$$
has the same solution of eq. (4.20) i.e. an oscillating RLC
superconducting circuit is equivalent to a normal
one with resonance frequency
$\Omega_0$ and resistance $R_{eff}=R_s \omega_0^2/\beta^2$.
 
For N superconducting wires or bodies the electric and magnetic part of the
lagrangian as well as eqs. (3.8) and (3.9) are the same as in the
ohmic case.
Without dissipation and superconducting kinetic contribution, 
the lagrangian in curved
spacetime is still given by eq. (3.7).
As far as the dissipation function is concerned, recalling that
only the normal currents dissipate and that they flow only near the
surface, we get
$$
\Phi = {1\over2}\sum_{a=1}^N 
R_{s\gamma_a}(t)I_{n\gamma_a}^2
\eqno (4.24)
$$
where
$$
R_{s\gamma_a}(t)={g_{ij}\over\sqrt{\ ^3\!g}}
\int_{\gamma_a} 
{dx^i t_{\gamma_a}^j\over\sigma_{\gamma_a}\eta_{\gamma_a}}
\eqno (4.25)
$$
in which $\eta_{\gamma_a}$ is the effective cross--section (4.14) of the
a--th wire. These equations allow us to write down the
generalization of the kinetic energy 
$$
W_{kin}(t)= {1\over 2\beta} \sum_{a=1}^N 
R_{s\gamma_a}(t)I^2_{s\gamma_a};
\eqno (4.26)
$$
and finally the lagrangian of a system of N superconducting wires or
bodies can be written as
$$
{\cal L} =  \sum_{a,b=1}^N \biggl[ {1\over 2}
\ L_{\gamma_a\gamma_b}(t) I_{\gamma_a} \ I_{\gamma_b} -
{1\over 2} 
\ {\widehat C_{\gamma_a\gamma_b}}(t) 
Q_{\gamma_a} Q_{\gamma_b}\bigg] + {1\over 2\beta} \sum_{a=1}^N 
R_{s\gamma_a}(t)I^2_{s\gamma_a}.
\eqno(4.27)
$$
The dissipation function is given by (4.24).
To get the Euler-Lagrange equations from lagrangian the (4.27) and the
dissipation function (4.24) we
have to know the relations between charge and current and between
charge and supercurrent. For a set of RLC circuits we have
$I_{\gamma_a}=\dot Q_{\gamma_a}$ and 
because of the general covariant form of eq. (4.2) (
$d/d\tau[\Lambda (j_s^\mu - j_s^\alpha u_\alpha u^\mu)]=F^\mu_\nu u^\nu$), 
the relation between $Q_{sa}$ and $Q_{na}$ is given in the curved 
case also by
$$
\dot Q_{s\gamma_a}=\beta Q_{n\gamma_a}.
\eqno (4.28)
$$
Using the method of Lagrange multipliers we get from (4.27) and (4.24)
$$
\sum_{b=1}^N \left [\ L_{\gamma_a\gamma_b}(t)\ddot Q_{\gamma_b}+
\dot L_{\gamma_a\gamma_b}(t)\dot Q_{\gamma_b} +
\widehat C_{\gamma_a\gamma_b}(t)\ Q_{\gamma_b}\right ]=
- R_{s\gamma_a}(t) \dot Q_{n\gamma_a}
\eqno(4.29)
$$
and by means of (4.28) we get the equations for the supercurrent 
$$
\eqalign {
\sum_{b=1}^N\left [\ L_{\gamma_a\gamma_b}(t)\dddot Q_{s\gamma_b}
+\right. &
(\beta L_{\gamma_a\gamma_b}(t)+\dot L_{\gamma_a\gamma_b}(t))
\ddot Q_{s\gamma_b} +
R_{s\gamma_a}(t) \ddot Q_{s\gamma_a}+
\cr
+&\widehat C_{\gamma_a\gamma_b}(t)\dot Q_{s\gamma_b} +\left.
\beta\widehat C_{\gamma_a\gamma_b}(t)\ Q_{s\gamma_b}\right ]= 0.
\cr }
\eqno(4.30)
$$
 
\bigskip
\noindent
{\bf 5. Conductors in the presence of matter}
\medskip
 
In this section we generalize our theory to the case in which 
the capacitor is filled by a dielectric and the inductance
has a magnetic kernel. Within the usual 
approximations the matter can be characterized
by a dielectric constant $\epsilon$ and a
magnetic permeability $\mu$ which enter into the constitutive equations
(see [8] p. 263):
$$
\eqalign {
H_{\alpha\beta}\ u^\alpha &= \epsilon F_{\alpha\beta}\ u^\alpha
\cr
\left ( g_{\alpha\beta}F_{\gamma\delta} +
g_{\alpha\gamma}F_{\delta\beta} +
g_{\alpha\delta}F_{\beta\gamma} \right )\ u^\alpha &=
\mu\ \left ( g_{\alpha\beta}H_{\gamma\delta} +
g_{\alpha\gamma}H_{\delta\beta} +
g_{\alpha\delta}H_{\beta\gamma} \right )\ u^\alpha
\cr}
\eqno(5.1)
$$
where $u^\alpha$ is the macroscopic 4--velocity of the medium.
With these definitions the Maxwell equations become
$$
\eqalign {
F_{\mu\nu,\sigma} + F_{\nu\sigma,\mu} &+ F_{\sigma\mu,\nu} = 0  
\cr
H^{\mu\nu}_{\ \ ;\nu} &= {4\pi\over c} j^\mu .
\cr}
\eqno(5.2)
$$
Under free--falling conditions $u^\alpha=\delta^\alpha_0$,
eqs. (5.1) reduce to 
$$
H_{0k}\ =\ \epsilon\ F_{0k} \qquad\qquad
F_{ik}\ =\ \mu\ H_{ik}.
\eqno(5.3)
$$
In this way the second pair of Maxwell equations becomes
$$
\eqalign {
F^{0\nu}_{\ \ ;\nu} &= {4\pi\over \epsilon}\ {j^0\over c}
\cr
F^{k\nu}_{\ \ ;\nu} &= {4\pi\over c}\ \mu\ j^k .
\cr }
\eqno(5.4)
$$
The action of the free electromagnetic field reads
$S_{em}=$ $(1/4\pi c)\int d^4x \sqrt{-g}\ H^{\mu\nu} F_{\mu\nu}$.
 
With these notations eqs. (3.18) and (3.22) become
(see [15] \S\S\ 30-33)
$$
\ ^0\!L_{\gamma_a\gamma_b}={\mu\over c^2}
\int_{\gamma_a}{}\int_{\gamma_b}{}\delta_{ij}
{dx^i_{\gamma_a}dx'^j_{\gamma_b}\over
|\bmit x_{\gamma_a}-\bmit x'_{\gamma_b}|},
\eqno(5.5)
$$
$$
\lambda^{ij}_{\gamma_a\gamma_b}=
{\mu\over c^2}
\int_{\gamma_a}{}\int_{\gamma_b}{}
{dx^i_{\gamma_a}dx'^j_{\gamma_b}\over
|\bmit x_{\gamma_a}-\bmit x'_{\gamma_b}|}+
{\mu\over 2c^2}
\int_{\gamma_a}{}\int_{\gamma_b}{\delta_{kl}}
\partial'^i\partial'^j
|\bmit x_{\gamma_a}-\bmit x'_{\gamma_b}|
dx^k_{\gamma_a}dx'^l_{\gamma_b},
\eqno(5.6)
$$
while (3.20) and (3.23) become
$$
Q^{\gamma_a} \ ^0\!\hat C_{\gamma_a\gamma_b}Q^{\gamma_b} = 
{1\over \epsilon}
\int_{\gamma_a}{}\int_{\gamma_b}{}
{dQ^{\gamma_a}\over dA_{\gamma_a}}
{dA_{\gamma_a}dA'_{\gamma_b}\over
|\bmit x_{\gamma_a}-\bmit x'_{\gamma_b}|}
{dQ^{\gamma_a}\over dA'_{\gamma_b}}.
\eqno(5.7)
$$
$$
Q^{\gamma_a}\chi^{ij}_{\gamma_a\gamma_b}Q^{\gamma_b}=
{1\over 2\epsilon}
\int_{\gamma_a}{}\int_{\gamma_b}{}
{dQ^{\gamma_a}\over dA_{\gamma_a}}
{dQ^{\gamma_b}\over dA'_{\gamma_b}}
\partial'^i\partial'^j
|\bmit x_{\gamma_a}-\bmit x'_{\gamma_b}|
dA_{\gamma_a}dA'_{\gamma_b}.
\eqno(5.8)
$$
Therefore the presence of matter with scalar dielectric or magnetic 
properties changes only the definition of the system parameters; they are
increased by $\epsilon$ and $\mu$ factors.
This is important because, in order to obtain great capacitances 
one can use suitable dielectrics: 
standard commercial types of capacitors can reach capacities of
about $10$ mF.
 
\bigskip
\noindent
{\bf 6. Application and Discussion}
\medskip
 
Let us now consider the device of Figure 1, located in the
$x-y$ plane with $x=x^1$ and $y=x^2$, and evaluate its response to
a periodic gravitational wave using realistic parameters. 
This circuit can be described as a set of six
conductors: two wires labelled $\gamma_1$ (the inductance
and resistance) and $\gamma_2$ (idealized connection wire) which 
connect four plane conductors labelled with $\gamma_3$, $\gamma_4$ (to 
form the capacitor $C_1$) and $\gamma_5$, $\gamma_6$ (the
capacitor $C_2$) with charge $Q_{\gamma_3}=-Q_1+Q$,
$Q_{\gamma_4}=Q_1-Q$, $Q_{\gamma_5}=Q_2+Q$ and $Q_{\gamma_6}=-Q_2-Q$;
$Q_1$ and $Q_2$ are the constant electrostatic charges on the plates of
the two capacitors in the absence of gravitational waves when there
is no current flow which satisfy
$$
\eqalign {
Q_1 + Q_2 =&Q_0,
\cr
Q_1/C_1 - Q_2/C_2 =&0.
\cr }
\eqno(6.1)
$$
The connection between the capacity of a condenser and the 
capacitance coefficients of the conductors $i$ and $j$ reads (see [15])
$$
C^{-1} \equiv\hat C_{ii}-2\hat C_{ij}+\hat C_{jj}.
\eqno (6.2)
$$
The lagrangian of this system can be written as
$$
{\cal L} = {1\over2} L(t)\dot Q^2 -
{1\over2} {(Q-Q_1)^2\over C_1(t)} -
{1\over2} {(Q+Q_2)^2\over C_2(t)}.
\eqno(6.3)
$$
Taking into account dissipation, the equation of the circuit becomes
$$
L(t)\ddot Q+\dot L(t)\dot Q+R(t)\dot Q+
\left ({1\over C_1(t)} + {1\over C_2(t)}\right ) Q =
{Q_1\over C_1(t)} - {Q_2\over C_2(t)}.
\eqno(6.4)
$$
Let us now consider the case of a weakly coupled gravitational--wave field.
Setting
$$
\eqalign {
\alpha_{+,\times}&=(\bmit e_{+,\times})_{ij}
{\lambda_{33}^{ij}\over L}
\cr
\kappa_{+,\times}&=-(\bmit e_{+,\times})_{ij}
(\chi_{11}^{ij}-2\chi_{12}^{ij}+\chi_{22}^{ij})\ C
\cr
\rho_{+,\times}&=(\bmit e_{+,\times})_{ij}
{\varrho_3^{ij}\over R},
\cr}
\eqno (6.5)
$$
we can separate the polarization and angular dependences of the interaction (figure
pattern). In this way we obtain
$$
\eqalign {
L(t)&= L[1+\epsilon_L(t)] =L[1+\alpha_+h_+(t)+\alpha_\times h_\times(t)],
\cr
{1\over C(t)}&= {1\over C} [1-\epsilon_C(t)] ={1\over C}
[1-\kappa_+h_+(t)-\kappa_\times h_\times(t)],
\cr
R(t)&= R[1+\epsilon_R(t)] =R[1+\rho_+h_+(t)+\rho_\times h_\times(t)].
\cr}
\eqno (6.6)
$$
For the sake of simplicity let us put
$$
\eqalign {
C_1 = a C\qquad&\qquad C_2 = C
\cr
\omega_0^2=&{a+1\over a}{1\over LC}.
\cr }
\eqno(6.7)
$$
With this definition of the factor $a$ we obtain
$$
Q_1={a\over a+1} Q_0\qquad\qquad
Q_2={1\over a+1} Q_0
\eqno(6.8)
$$
and the equation of the circuit can be written as
$$
\eqalign {
(1 + \epsilon_L(t))\ddot Q +
2\gamma
\left (1 + {\dot\epsilon_L(t)\over 2\gamma} + \epsilon_R(t) \right )
\dot Q +
\omega_0^2&
\left (1-{\epsilon_{C_1}(t) + a\epsilon_{C_2}(t)\over a+1} \right ) Q =
\cr
=\omega_0^2 Q_0
{a(\epsilon_{C_1}(t) - \epsilon_{C_2}(t))\over (a + 1)^2}.&
\cr }
\eqno(6.9)
$$
As the time dependent coefficients in
the l.h.s. are very small compared with unity
their unique effect will be an amplitude and frequency modulation of the
unperturbed charge motion
$$
\ddot Q + 2\gamma \dot Q + \omega_0^2 Q =
v_+ h_+(t) + v_\times h_\times(t)
\eqno(6.10)
$$
where
$$
v_{+,\times} = \omega_0^2 Q_0
{a (\kappa_{1,+,\times} - \kappa_{2,+,\times})\over (a + 1)^2}.
\eqno(6.11)
$$
In equation (6.10) the interaction between the gravitational wave 
and the circuit is non--parametric. The dropped modulations are so small that
they cannot be observed being a second order effect with respect to the
solution of (6.10). If we consider a periodic gravitational wave
$h_{+,\times}(t)=A_{+,\times}\ \cos{(\omega_g t+\phi_{+,\times})}$,
then the solution of (6.10) can be written as
$$
Q(t)=\ Q_+(t) + Q_\times (t)
\eqno(6.12)
$$
$$
Q_{+,\times}(t)=
{a(\kappa_{1,+,\times}-\kappa_{2,+,\times})\over (a+1)^2}
{2 A_{+,\times}Q_0{\cal Q}\over
\sqrt{1+{(\omega_0^2-\omega_g^2)^2\over4\gamma^2\omega_g^2}}}
{\omega_0\over\omega_g}\ \sin{(\omega_gt+\phi_{+,\times}+\delta)}
$$
where ${\cal Q}=\omega_0/4\gamma$ is the quality factor of the 
circuit and 
$$
\delta=\arctan{{\omega_0^2-\omega_g^2\over2\gamma\omega_g}}.
\eqno(6.13)
$$
In this system the effect of the gravitational wave is to cause a 
current $I(t)$ to flow in the circuit with
$$
I(t)={a(\kappa_1-\kappa_2)\over (a+1)^2}
{2 A\omega_0 Q_0{\cal Q}\over
\sqrt{1+{(\omega_0^2-\omega_g^2)^2\over4\gamma^2\omega_g^2}}}
\ \cos{(\omega_gt+\phi+\delta)}.
\eqno(6.14)
$$
Near resonance (i.e. $(\omega_0-\omega_g)^2<<(2\gamma)^2$) we have
$$
I_{near\ res}(t)={a(\kappa_1-\kappa_2)\over (a+1)^2}
{2 A\omega_0 Q_0 {\cal Q}\over
1+2\left ({\omega_0-\omega_g\over2\gamma}\right )^2}
\ \cos{(\omega_gt+\phi+\delta_{near\ res})}
\eqno(6.15)
$$
where $\delta_{near\ res}\simeq{\omega_0-\omega_g\over\gamma}$.
This device behaves formally like a Weber bar with the initial 
static charge $Q_0$ corresponding to the rest length of the bar.
The device has two advantages:
firstly, it can be easily tuned to receive radiation from periodic
sources like pulsars and binary stars ($\nu_g\ >$ few Hz); and secondly,
the measurement is a null measurement (the expected current
versus zero).
 
Let us now consider the superconducting case.
Using the same notations eq. (6.4) becomes
$$
L(t)\ddot Q+\dot L(t)\dot Q+R_s(t)\dot Q_n+
\left ({1\over C_1(t)} + {1\over C_2(t)}\right ) Q =
{Q_1\over C_1(t)} - {Q_2\over C_2(t)}.
\eqno(6.16)
$$
For weak gravitational waves we have 
$$
\eqalign {
(1+\epsilon_L(t))\ddot Q + \dot \epsilon_L(t)\dot Q +
2\gamma_s [1 + \epsilon_R(t)]\dot Q_n &+
\omega_0^2
\left (1-{\epsilon_{C_1}(t) + a\epsilon_{C_2}(t)\over a+1} \right ) Q =
\cr
=\omega_0^2 Q_0
{a(\epsilon_{C_1}(t) - \epsilon_{C_2}(t))\over (a + 1)^2}.&
\cr }
\eqno(6.17)
$$
Again the time dependent coefficients in the l.h.s. of
equation (6.17) produce only a very small frequency and amplitude
modulation of the solution of the following equation
$$
\ddot Q + 2\gamma_s \dot Q_n + \omega_0^2 Q =
v_+ h_+(t) + v_\times h_\times(t)
\eqno(6.18)
$$
where
$$
v_{+,\times} = \Omega_0^2(1+2\Gamma) Q_0
{a (\kappa_{1,+,\times} - \kappa_{2,+,\times})\over (a + 1)^2}.
\eqno(6.19)
$$
In equation (6.18) the gravitational--wave electromagnetic--circuit interaction
is purely non--parametric.
If we use relation (4.28) in (6.18) we find the equation of the
supercurrent as
$$
\dddot Q_s + \beta (1+2\Gamma) \ddot Q_s +
\Omega_0^2(1+2\Gamma)\dot Q_s +
\beta\Omega_0^2(1+2\Gamma) Q_s =
\beta [v_+ h_+(t) + v_\times h_\times (t)].
\eqno(6.20)
$$
If
$h_{+,\times}(t)=A_{+,\times}\ \cos{(\omega_g t+\phi_{+,\times})}$,
then the solution of (6.20) can be written as
$$
Q_s(t)=Q_{s,+}(t) + Q_{s,\times}(t)
\eqno(6.21)
$$
$$
Q_{s,+,\times}(t)=
{a(\kappa_{1,+,\times}-\kappa_{2,+,\times})\over (a+1)^2}
{2 A_{+,\times}(1+2\Gamma)Q_0{\cal Q}_{eff}\over T(\omega_g)}
{\Omega_0\over\omega_g}\ \sin{(\omega_gt+\phi_{+,\times}+\delta)}
$$
where
$$
\tan{\delta}=
{(1+2\Gamma)(\Omega_0^2-\omega_g^2)\over 2\gamma_0\omega_g
\left (1+{\Omega_0^2-\omega_g^2\over 2\gamma_0\beta}\right )},
\eqno(6.22)
$$
$$
T(\omega_g)=
\left (1+{\Omega_0^2-\omega_g^2\over 2\gamma_0\beta}\right )
\sqrt{1+\tan^2{\delta}}
\eqno(6.23)
$$
and
$$
{\cal Q}_{eff}={\Omega_0\over4\gamma_0}={\beta\over 4\Gamma\Omega_0}=
{\beta^2\over 2 \Omega_0} {L\over R_s}
\eqno(6.24)
$$
is the quality factor of the superconducting circuit which is
several orders of magnitude greater than that of the ohmic circuit.
In this way the gravitational wave gives rise to a supercurrent
$$
I_s(t)= {a(\kappa_1-\kappa_2)\over (a+1)^2}
{2 A(1+2\Gamma)\Omega_0Q_0{\cal Q}_{eff}\over T(\omega_g)}
\ \cos{(\omega_gt+\phi+\delta)}
\eqno(6.25)
$$
in which, for the sake of simplicity, we have considered only one
polarization state and dropped the indices $+,\times$.
Near resonance (i. e. $(\Omega_0^2-\omega_g^2)^2<<(2\gamma_0)^2$)
we have
$$
I_{s\ near\ res}(t)={a(\kappa_1-\kappa_2)\over (a+1)^2}
{2 A\Omega_0 Q_0 {\cal Q}_{eff}\over
1+2\left ({\Omega_0-\omega_g\over2\gamma_0}\right )^2}
\ \cos{(\omega_gt+\phi+\delta_{near\ res})}
\eqno(6.26)
$$
where $\delta_{near\ res}\simeq{\Omega_0-\omega_g\over\gamma_0}$.
To get the total current we have to use (4.28) for the normal
current,
$$
I_n(t)=- {\omega_g\over \beta}
{a(\kappa_1-\kappa_2)\over (a+1)^2}
{2 A(1+2\Gamma)\Omega_0Q_0{\cal Q}_{eff}\over T(\omega_g)}
\ \sin{(\omega_gt+\phi+\delta)}.
\eqno(6.27)
$$
We see that this current (in agreement with (4.4) and (4.5)) is
smaller by a factor $\omega_g/\beta$ and it is
$\pi/2$ out of phase with respect to the supercurrent.
The response of a superconducting
device differs from the response in the ohmic case because of
different quality factors.
 
Let us now calculate explicitly the parameters of the circuit.
The eqs. (3.22-24) and (3.18-20) or (5.5-6) and (5.7-8) give:
$$
\eqalign {
\delta_{ij} \lambda^{ij}_{\gamma_a\gamma_b} &= 2 L_{\gamma_a\gamma_b}
\cr
\delta_{ij} Q^{\gamma_a} \chi_{\gamma_a\gamma_b}^{ij} Q^{\gamma_b} =
Q^{\gamma_a} \hat C_{\gamma_a\gamma_b} Q^{\gamma_b}&
\quad \Rightarrow \quad 
\delta_{ij} \chi_{\gamma_a\gamma_b}^{ij} = \hat C_{\gamma_a\gamma_b}
\cr
\delta_{ij} \rho^{ij}_{\gamma_a} &= R_{\gamma_a}
\cr }
\eqno (6.28)
$$
We note that
$$
\partial'^1\partial'^1
|\bmit x_{\gamma_a}-\bmit x'_{\gamma_b}|=
{(y-y')^2+(z-z')^2\over |\bmit x_{\gamma_a}-\bmit x'_{\gamma_b}|^3} , etc.
\eqno(6.29)
$$
so that 
$$
Q^{\gamma_a}\chi^{ii}_{\gamma_a\gamma_b} Q^{\gamma_b} = 
{1\over 2} Q^{\gamma_a} \hat C_{\gamma_a\gamma_b} Q^{\gamma_b} -
{1\over 2\epsilon} 
\int_{\gamma_a}{}\int_{\gamma_b}{}
{dQ^{\gamma_a}\over dA_{\gamma_a}}
{dQ^{\gamma_b}\over dA'_{\gamma_b}}
{(x^i_{\gamma_a}-x'^i_{\gamma_b})^2\over
|\bmit x_{\gamma_a}-\bmit x'_{\gamma_b}|^3}
dA_{\gamma_a}dA'_{\gamma_b}
\eqno(6.30)
$$
(no summation over i).
 
Let us now consider the case of a capacitor made of two circular
plates set perpendicular to the $x$ axis and at a distance $d$ 
apart.
If $\gamma_a = \gamma_b$ then
$$ 
\chi^{11}_{\gamma_a\gamma_a}  =
{1\over 2}  \hat C_{\gamma_a\gamma_a}
\eqno (6.31)
$$
and
$$
\chi^{22}_{\gamma_a\gamma_a} = \chi^{33}_{\gamma_a\gamma_a} =
{1\over 4} \hat C_{\gamma_a\gamma_a}.
\eqno(6.32)
$$
If $\gamma_a \not= \gamma_b$, assuming that $d$ is much 
smaller than the
diameter of the circular plates, by means of (6.28) and the
symmetry of the problem, we approximately get
$$
\eqalign {
\chi^{11}_{\gamma_a\gamma_b}&=  \hat C_{\gamma_a\gamma_b} - 
{1\over 2} \hat C_{\gamma_a\gamma_a} 
\cr
\chi^{22}_{\gamma_a\gamma_b} = \chi^{33}_{\gamma_a\gamma_b}&=
{1\over 4} \hat C_{\gamma_a\gamma_a}
\cr}
\eqno(6.33)
$$
which is valid also for $\gamma_a = \gamma_b$ as can be seen by 
comparing (6.33) with (6.31--32).
Similarly one can show that $\chi^{12}_{\gamma_a\gamma_b}=0$.
 
Now let us assume that the wave comes from the $z=x^3$ axis perpendicular 
to the plane containing our circuit (see Figure 1). In this case
we have for the polarization tensors
$$
(e_+)_{\mu\nu} =
\left(
\matrix {0 & 0 & 0 & 0 \cr
0 & 1 & 0 & 0 \cr
0 & 0 & -1 & 0 \cr
0 & 0 & 0 & 0}
\right)
\eqno(6.34)
$$
$$
(e_\times)_{\mu\nu} =
\left(
\matrix {0 & 0 & 0 & 0 \cr
0 & 0 & 1 & 0 \cr
0 & 1 & 0 & 0 \cr
0 & 0 & 0 & 0}
\right)
\eqno(6.35)
$$
With reference to eq. (6.5) let us calculate $\kappa_+$; one finds
$$
\eqalign {
\kappa_+ &= - (e_+)_{ij} \left (
\chi_{11}^{ij} - 2 \chi_{12}^{ij} + \chi_{22}^{ij}
\right ) C =
\cr
&=\left [ 
\left ( \chi_{11}^{22} - \chi_{11}^{11} \right ) -
2 \left ( \chi_{12}^{22} - \chi_{12}^{11} \right ) +
\left ( \chi_{22}^{22} - \chi_{22}^{11} \right )
\right ] C.
\cr}
\eqno (6.36)
$$
Taking the $x$ axis parallel to the plates of $C_1$ and the y axis parallel
to the plates of $C_2$ we obtain
$$
\eqalign {
\left (\kappa_+\right )_1 &= 1, \qquad\qquad
\left (\kappa_\times\right )_1 = 0
\cr
\left (\kappa_+\right )_2 &= -1, \qquad\qquad
\left (\kappa_\times\right )_2 = 0
\cr }
\eqno (6.37)
$$
Substituting (6.37) in (6.15) and (6.26), and taking $a=1$ one gets
$$
\eqalign {
I_{res}&= A_+ \omega_0 Q_0 {\cal Q} 
\cos {(\omega_0 t + \phi)},
\cr
I_{res.s}&= A_+ \Omega_0 Q_0 {\cal Q}_{eff} 
\cos {(\omega_0 t + \phi)}.
\cr }
\eqno (6.38)
$$
We now estimate the minimal gravitational amplitude which can be 
detected by our device.  Once the device is isolated from the outside world
by means for instance of mechanical filters, as those used in gravitational 
bar detectors and Faraday cages, the main noise source is the thermal one,
as well for the mechanical part of the detector as for the electric part, e.g.
the back--action noise of the amplifier, at cryogenic temperature and for the
best amplifiers, is at most of the order of magnitude of the thermal noise.
In the following we shall derive formulae for the thermal noise in the
electric part of the detector. The thermal spectral noise in the mechanical
part ($2 \delta l/l = h$, see equation (2.17)) is given by
$\sqrt{16k_BT/M \tau l^2 {\omega}^4}$
where $T$, $M$, $\tau$, and $l$ denote temperature, effective mass,
life-time of the fundamental mode, and effective length of the detector, $k_B$
is the Boltzmann constant and $\omega$ denotes the angular frequency of the
gravitational wave, e.g. see [24]. Assuming the numbers $T=4$ K, $M=10^4$ kg,
$\tau = 10$ sec, $l=10$ m, and $\omega /2 \pi = 60$ Hz yields about
$2.1 \times 10^{-23}$ for $h$, measured through a period of 4 months. This
number is as small as the smallest number in the equation (6.45) below, i.e.
for the superconducting circuits in question the mechanical part has to be
adjusted to those numbers.
 
We now make the hypothesis that the only resistance in our circuit is that of
our scheme in Figure 1. The mean square fluctuation of the voltage noise
is therefore given for $ t_{obs} >> {\cal Q}  / {\omega}_0 $ by ([25] p.288)
$$
<V_n^2>= {k_BT\over C_{eff}}
\eqno (6.39)
$$
where $C_{eff}$ is the total capacitance of the capacitors. The mean noise
energy is
$$
C_{eff}<V_n^2>= k_BT
\eqno (6.40)
$$ 
The energy dissipated by the ohmic and superconducting current 
produced by the gravitational wave is
$$
\eqalign {
W &=\int{R I^2 dt}={1\over 2} R I_0^2 t_{obs}
\cr
W_s &=\int{R_s I_n^2 dt}={1\over 2} R_s I_{n0}^2 t_{obs}
\cr }
\eqno (6.41)
$$
where $t_{obs}$ is the observational time and $I_0$ the amplitude 
of the current $I$.  
Equating (6.40) and (6.41) one finds the minimum detectable
current after an observation time $t_{obs}$,
$$
\eqalign {
I_0&= \sqrt {{2k_B\ T\over R\ t_{obs}}}
\cr
I_{n0}&= \sqrt {{2k_B\ T\over R_s\ t_{obs}}}
\cr }
\eqno (6.42)
$$
At resonance this implies (by means of eqs. (6.7), (6.15) and
${\cal Q} = \omega_0 L/(2 R)$ for the ohmic circuit and (4.22), 
(6.24) and (6.27) for the superconducting one) that
$$
\eqalign {
A_{noise}&= {1\over V_0}\ 
\sqrt{{(a+1)^4\over a^2 (\kappa_1-\kappa_2)^2}
{k_BT \over {\cal Q} C \omega_0  t_{obs}}}
\cr
A_{noise\ s}&=  {1\over V_0}\ 
\sqrt{{(a+1)^4\over a^2 (\kappa_1-\kappa_2)^2}
{k_BT \over {\cal Q}_{eff} C \omega_0  t_{obs}}}
\cr }
\eqno (6.43)
$$
where $V_0=Q_0/C$. If $a=\kappa_1=-\kappa_2=1$ then
$$
\eqalign {
A_{noise}&= {2\over V_0}\ 
\sqrt{{k_BT \over {\cal Q} C \omega_0  t_{obs}}}
\cr
A_{noise\ s}&= {2\over V_0}\ 
\sqrt{{k_BT \over {\cal Q}_{eff} C \omega_0  t_{obs}}}
\cr }
\eqno (6.44)
$$
Taking $V_0=10^5$ Volt, $T=4$ K,
$C=10^{-2}$ F,
$\omega_0=2\pi\times 60$ rad/sec,
$t_{obs}= 4$ months = $10^7$ sec, ${\cal Q}=10^3$ and
${\cal Q}_{eff}=10^6$ one has 
$$
A_{noise}\simeq 7.6 \times 10^{-22},\qquad\qquad
A_{noise\ s}\simeq 2.4 \times 10^{-23}
\eqno (6.45)
$$ 
Finally we compute the order of magnitude of the normal and 
superconducting current, produced by a gravitational wave with 
amplitudes $A_+=2\times 10^{-21}$ and $A_+=5\times 10^{-23}$ which 
correspond respectively to a signal to noise ratio equal to 2; 
we find
$$
\eqalign {
I_n &\simeq\ 7.5 \times 10^{-13}\ A
\cr
I_s &\simeq\ 2 \times 10^{-11}\ A
\cr}
\eqno(6.46)
$$
that are within the possibilities of actual measurements with SQUIDs.
If we go down with the temperature in the ultracryogenic zone 
($\sim 40$ mK) we obtain
$$
\eqalign {
A_{noise}&\simeq 7.6 \times 10^{-23}\qquad\qquad
I\simeq 7.5 \times 10^{-14}\ A 
\cr
A_{noise\ s}\simeq &2.4\times 10^{-24}\qquad\qquad 
I_s\simeq 2 \times 10^{-12}\ A 
\cr }
\eqno (6.47)
$$
Further improvements can come either from  different arrangements of the 
circuit elements (for instance one can consider a series of many 
capacitors). As far as the 
last question is concerned, today's limitations in the superconducting 
case are due to the energy dissipation in the material which acts as a 
mounting of the circuit.
For this reason we have set ${\cal Q}_{eff}=10^6$ but there 
are no theoretical limitations in thinking of quality factors higher by 
several orders of magnitude: it is in fact only a technological 
problem.
We think that a serious research in this field would be very useful.
 
It is to be expected that when the device is operating near the mechanical
resonance the sensitivity should increase. However this situation deserves
a more detailed investigation.
 
Recall now that we have calculated the currents in TT--gauge. In 
actual experiments we measure currents in the laboratory reference 
frame (FNC). Because of the charge conservation we can write 
$$
I^{FNC}\ dy^0\ =\ I^{TT}\ dx^0
\eqno(6.48)
$$
where $y^\mu$ are the FNC coordinates while $x^\mu$ are the 
TT coordinates. As
$x^0=y^0-{1\over 4}h_{ij}^{\ \ ,0}y^iy^j$ we have 
$$
dx^0=
\left ( 1-{k^2\over 4} h_{ij}^{(2)} y^iy^j \right ) dy^0 -
{k\over 2}h_{ij}^{(1)} y^i dy^j
\eqno(6.49)
$$
where the superscript $(i)$ is the i--th derivative of $h$ with 
respect to its argument.
Therefore the relation between the current in the two gauges reads
$$
I^{FNC} = I^{TT} 
\left ( 1-{k^2\over 4} h_{ij}^{(2)} y^iy^j -
{k\over 2}h_{ij}^{(1)} y^i {dy^j\over dy^0} \right ).
\eqno(6.50)
$$
In our approximation both first order corrections are negligible and so
$$
I^{FNC}\ =\ I^{TT}
\eqno(6.51)
$$
 
We point out that our theoretical limits in the
magnitude of the gravitational--wave amplitude detectable from periodic
sources like the Crab pulsar are better than those experimentally obtained
by a Japanese research group ($A<5\times 10^{-22}$ [20]; see also [21]).
Therefore, a superconducting device at liquid helium temperature could
give new limits for the emission of gravitational waves from this pulsar.
As to vary an electromagnetic frequency is much simpler than varying a
mechanical one, our detectors can be easily adjusted to any
gravitational--wave frequency measurable on Earth.

\vfill\eject
\vskip 1.2truecm
\noindent
{\bf References}
\bigskip
 
\ref {1.~~~G. A. Lupanov, Sov. Phys. JETP {\bf 25}, 76 (1967);
F. Cooperstock, Ann. Phys. (N.Y.) {\bf 47}, 173 (1968);
D. Boccaletti, V. De Sabbata, P. Fortini and C. Gualdi, Nuovo Cimento
{\bf B 70}, 129 (1970); P. Fortini, C. Gualdi and A. Ortolan, Nuovo 
Cimento {\bf B 106}, 395 (1991).}
 
\ref {2.~~~K. S. Thorne in {\it Three hundred years of gravitation};
eds. S. Hawking and W. Israel, Cambridge Univ. Press, Cambridge (1987).}
 
\ref {3.~~~V. B. Braginsky, M. B. Menskii Sov. Phys. JEPT Lett. {\bf
13}, 417 (1971); V. B. Braginsky, L. P. Grishchuk, A. G. Doroskevich,
Ya. B. Zel'dovich, I. D. Novikov and M. V. Sazhin Sov. Phys. JETP
{\bf 38}, 865 (1973); F. Pegoraro, E. Picasso and L. A. Radicati
J. of Phys. {\bf A 11}, 1949 (1978); C. M. Caves Phys. Lett. {\bf
B 80}, 393 (1979).}
 
\ref {4.~~~C. W. Misner, K. S. Thorne and J. A. Wheeler
{\it Gravitation}; Freeman, San Francisco (1973).}
 
\ref {5.~~~P. Fortini and C. Gualdi, Nuovo Cimento {\bf B 71}, 37 (1982).}
 
\ref {6.~~~G. Callegari, P. Fortini and C. Gualdi, Nuovo Cimento
{\bf B 100}, 421 (1987).}
 
\ref {7.~~~S. Weinberg, {\it Gravitation and Cosmology}; 
Wiley, New York (1972) Ch. 5.}
 
\ref {8.~~~R. C. Tolman, {\it Relativity, Thermodynamics and Cosmology};
Clarendon Press, Oxford (1966).}
 
\ref {9.~~~V. N. Golubenkov and Ya. A. Smorodinsky, Zh. eksper. teor. Fiz.
{\bf 31}, 330 (1956).}
 
\ref {10.~~~G. Sch\"afer and H. Dehnen, J. Phys. {\bf A 13}, 2703 (1980).}
 
\ref {11.~~~B. Mours and M. Yvert, Phys. Lett. {\bf A 136}, 209 (1989).}
 
\ref {12.~~~F. London, Superfluids, Dover Publ. (1961); F. London Phys. 
Rev. {\bf 51}, 678(L) (1937).}
 
\ref {13.~~~H. Goldstein, {\it Classical Mechanics}, second edition, 
Addison Wesley (1980).}
 
\ref {14.~~~L. D. Landau, E. M. Lifshitz, {\it Course of Theoretical
Physics Vol. 2: The Classical Theory of Fields}, Pergamon Press (1975).}
 
\ref {15.~~~L. D. Landau, E. M. Lifshitz, {\it Course of Theoretical
Physics Vol. 8: Electrodynamics of Continuous Media}, Pergamon Press 
(1960).}
 
\ref {16.~~~R. Schrader Phys. Lett. {\bf B 143}, 421 (1984).}
 
\ref {17.~~~J. L. Stone and W. H. Hartwig J. Appl. Phys. {\bf 39},
2665 (1968).}
 
\ref {18.~~~McLachlan, {\it Theory and Application of Mathieu 
Functions}, Clarendon Press (1947).}
 
\ref {19.~~~L. D. Landau, E. M. Lifshitz, {\it Course of Theoretical
Physics Vol. 1: Mechanics}, Pergamon Press (1960).}
 
\ref {20.~~~S. Owa, M. K.  Fujimoto, H. Hirakawa, K. Morimoto, 
T. Suzuki, K. Tsubono, Proc. IV Marcel Grossmann meeting on Gen.Rel. 
p. 571-8, Hu Ning Ed. North Holl. (1986).}
 
\ref {21.~~~H. Hirakawa, K. Narihara Phys. Rev. Lett. {\bf 35}, 
330 (1975);
H. Hirakawa, K. Narihara, M. K. Fujimoto J. Phys. Soc. Jap. {\bf 41}, 
1093 (1976);
T. Suzuki, K. Tsubono, H. Hirakawa Phys. Lett. {\bf A 67}, 2 (1978);
K. Oide, H. Hirakawa Phys. Rev. {\bf D 20}, 2480 (1979);
S. Kimura, T. Suzuki, H. Hirakawa Phys. Lett. {\bf A 81}, 302 (1981);  
M. K. Fujimoto Proc. III Marcel Grossmann meeting on Gen.Rel. 
p. 1447, Hu Ning Ed. North Holl.(1983);
K. Kuroda, H. Hirakawa Japan. J. Phys. {\bf 22}, 1005 (1983);
K. Kuroda, K. Tsubono, H. Hirakawa Jap. Jour. Appl. Phys. {\bf 23}, 
L415 (1984);
K. Tsubono, M. Ohashi, H. Hirakawa Jap. Jour. Appl. Phys. {\bf 25}, 
622 (1986);
Y. Nagashima, S. Owa, K. Tsubono, H. Hirakawa Rev. Sci. Instrum. 
{\bf 59}, 112-4 (1988).}
  
\ref {22.~~~K. S. Thorne in {\it Gravitational Radiation}; eds. N. Deruelle
and T. Piran, North--Holland Publishing, Amsterdam (1983) p. 1.}
 
\ref {23.~~~L. Baroni, G. Callegari, P. Fortini, C. Gualdi, 
M. Orlandini in {\it Proceedings of the Fourth Marcel Grossmann 
Meeting on General Relativity}; ed. R. Ruffini (Elsevier Science 
Publ., Amsterdam 1986) p. 641.}
 
\ref {24.~~~J. Colas, L. Massonnet, B. Mours and M. Yvert, {\it D\'etecteur
d'ondes gravitationnelles capacitif et bruit thermique},
LAPP--EXP--90--01, May 1990.}
 
\ref {25.~~~A. Papoulis, {\it Probability, random variables and
stochastic processes} - Second Edition - McGraw--Hill, New York
(1984).}

\vskip 1.2truecm
\noindent
{\bf Figures}
\bigskip

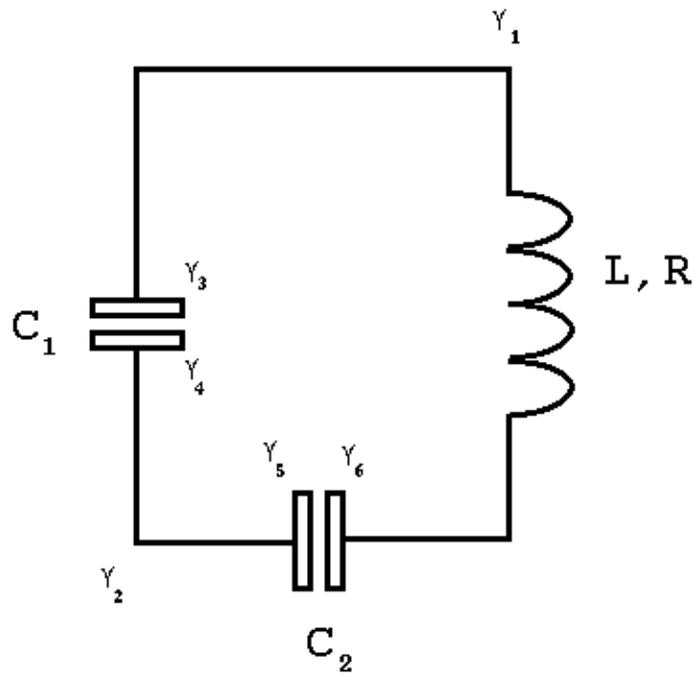
Fig.1 --- A scheme of the RLC circuit describrd in the text which can 
be used as a detector of gravitational waves. A static charge is 
distributed on the plates of the condensers $C_1$ and $C_2$.

\bye